\documentstyle[11pt,graphicx]{article}
\title{New Primordial-Magnetic-Field Limit from The Latest LIGO S5 data}

\author{S. Wang$^{1,2}$\cite{email},  \\
   \small $^1$Department of Modern Physics, University of Science and Technology of China, Hefei, Anhui 230026, China. \\
   \small $^2$Institute of Theoretical Physics, Chinese Academy of Science, Beijing 100080, China.}
\date{}

\begin{document}
\maketitle
\baselineskip=19truept

\def\vek{\vec{k}}
\renewcommand{\arraystretch}{1.5}
\newcommand{\be}{\begin{equation}}
\newcommand{\ee}{\end{equation}}
\newcommand{\ba}{\begin{eqnarray}}
\newcommand{\ea}{\end{eqnarray}}

\sf

\begin{center}
\Large  Abstract

\end{center}

\baselineskip=19truept

\begin{quote}

Since the energy momentum tensor of a magnetic field always contains a spin-2 component in its anisotropic stress,
stochastic primordial magnetic field (PMF) in the early universe must generate stochastic gravitational wave (GW) background.
This process will greatly affect the relic gravitational wave (RGW),
which is one of major scientific goals of the laser interferometer GW detections.
Recently, the fifth science (S5) run of laser interferometer gravitational-wave observatory (LIGO)
gave a latest upper limit $\Omega_{GW}<6.9\times10^{-6}$ on the RGW background.
Utilizing this upper limit, we derive new PMF Limits:
for a scale of galactic cluster $\lambda=1$ Mpc, the amplitude of PMF, that produced by the electroweak phase transition (EPT),
has to be weaker than $B_{\lambda} \leq 4\times 10^{-7}$ Gauss;
for a scale of supercluster $\lambda=100$ Mpc, the amplitude of PMF has to be weaker than $B_{\lambda} \leq 9\times 10^{-11}$ Gauss.
In this manner, GW observation has potential to make interesting contributions to the study of primordial magnetic field.

\end{quote}

PACS numbers: 95.85.Sz, 04.30.Tv, 04.80.Nn

\newpage

\begin{center}
{\bf 1. Introduction}
\end{center}

Although the magnetic fields in galaxies and galactic clusters have been observed for many years \cite{Kronberg},
the origin of these magnetic fields still remains a mystery.
One of the most promising candidate origins is the primordial magnetic field (PMF) produced in the early universe \cite{Widrow}.
Such a seed field can be produced by the inflation \cite{Turner,Ratra,Ohta},
or by the electroweak phase transition (EPT) \cite{Baym,Sigl,KKS}.
In the past, PMF has been constrained mainly by using their various effects
on cosmic microwave background (CMB) anisotropies and polarization \cite{Kosowsky1,Giovannini1,Barrow,Durrer1,Kosowsky2,KLR}.
Based on the Faraday rotation of CMB linear polarization caused by PMF,
Kahniashvili et al \cite{Kahniashvili} utilized the 5-year WMAP (WMAP5) $B$-mode polarization limit
to give an upper limit on the amplitude of PMF:
$B_{\lambda}\le 6\times 10^{-8}$ Gauss on a scale $\lambda= 1$ Mpc.
Besides, Kristiansen and Ferreira \cite{Kristiansen} forecasted the constraints on PMF from future CMB polarization experiments.
On the other hand, a stochastic PMF itself can generate the relic gravitational wave (RGW) background \cite{Mack,Tsagas}.
This is because the energy momentum tensor of a magnetic field always contains a spin-2 component in its anisotropic stress.
Utilizing the nucleosynthesis limit on gravitational wave (GW),
Caprini and Durrer \cite{Durrer2} gave extraordinarily strong constraints on the amplitude of PMF:
for PMF produced by EPT, $B_{\lambda}\le 10^{-27}$ Gauss; and for PMF produced by inflation, $B_{\lambda}\le 10^{-39}$ Gauss.
In addition, Giovannini and Kunze \cite{Giovannini2} gave a comprehensive treatment of the PMF's effects on CMB maps,
and obtained a constraint $B_{\lambda}\le 5\times 10^{-9}$ Gauss by using the WMAP5 $B$-mode polarization limit.

Laser interferometer GW detection is one of the most important approaches to detect the RGW.
Using LIGO S4 data and cross-correlation technique \cite{Allen},
LIGO scientific collaboration \cite{Abbott1} gave an upper limit $\Omega_{GW}<6.5\times10^{-5}$ on the total energy density of RGW.
In addition, LIGO scientific collaboration and ALLEGRO collaboration \cite{Abbott2}
also gave an upper limit $\Omega_{gw}(\nu)\leq 1.02$ on the stochastic GW background.
Especially, in a latest work, using LIGO S5 data,
LIGO scientific collaboration and Virgo collaboration \cite{ligo} gave a latest upper limit $\Omega_{GW}<6.9\times10^{-6}$,
which improves LIGO S4 result for one order of magnitude.
So in this work, we shall give the constraints on PMF by using the latest LIGO S5 result.
For the PMF, we shall focus on that produced by EPT.
As seen below, it yields a level spectrum of RGW in the range $(10^{-2}-10^{4})$ Hz, which covers the bands of operation of LIGO.
PMF can also be produced by inflation.
However, since the PMF produced by inflation affects the RGW
only in very high frequency range $v \geq 10^{9}$ Hz that is out of the bands of operation of LIGO,
we will not discuss this class of PMF.
Our work differs from previous papers in the following two aspects:
First, in previous papers \cite{Durrer2,Giovannini2},
PMF is directly treated as the source of RGW. However, RGW has already existed before the EPT.
That is to say, if one treated the PMF produced by EPT as the source of RGW,
a lot of important information about ultra-early universe, such as inflation and reheating, will be lost.
Therefore, for a complete treatment,
the PMF produced by EPT should only be viewed as a subsequent factor that affect the RGW, which is adopted in our work.
Second, in previous papers,
only the nucleosynthesis limit \cite{Durrer2} and the WMAP5 $B$-mode polarization limit \cite{Giovannini2}
have been used to constrain the RGW and to give the constraint on PMF.
Since the laser interferometer GW detection is also a crucial approach to detect RGW,
it would be very interesting to give a constraint on PMF by using the data of current GW detection.
To our present knowledge, this issue has not been investigated before.
So in this work, we shall give new limits on the amplitude of PMF by using the latest LIGO S5 data.

The organization of this paper is as follows.
In section 2, from the inflationary stage up to the accelerating stage,
the scale factor $a(\tau)$ is specified by the continuity conditions for the subsequential stages of expansion.
In section 3, the analytical solution of RGW is determined with the coefficients being fixed
by continuity condition joining two consecutive expansion stages.
In particular, the important impact of PMF produced by EPT is included into our calculation.
In section 4, we calculate the spectrum $h(\nu,\tau_0)$ and the one-sided spectral density $S_h(\nu)$ of RGW,
and then give the theoretical predictions of RGW, which can be directly compared with the latest LIGO data.
In section 5, by comparing the theoretical predictions of RGW with LIGO S5 data, we give the constraints on the amplitude of PMF.
Several technical derivations are deferred to two appendices.
It should be noted that,
our conventions are different from that of Ref. \cite{Durrer2}:
the scale factor $a(\tau)$ has the unit of cm,
while the conformal time $\tau$ and the co-moving distance $x$ do not have units.
Besides, Greek indices run from $0$ to $3$, Latin indices run from $1$ to $3$,
the subscript ``0'' always indicates the present value of the corresponding quantity, and the unit with $c=\hbar=k_B=1$ is used.

\

\begin{center}
{\bf 2. Expansion History of The Universe}
\end{center}

From the inflationary stage up to the currently accelerating stage,
the
expansion of the Universe can be described by the
spatially flat Robertson-Walker spacetime with a metric
\be
ds^2=a^2(\tau)[-d\tau^2+\delta_{ij}dx^idx^j].
\ee
The scale factor $a(\tau)$
for the successive stages can be approximately
described by the following forms
\cite{grishchuk1}:

The inflationary stage:
\be \label{inflation}
a(\tau)=l_1|\tau|^{1+\beta},\,\,\,\,-\infty<\tau\leq \tau_1,
\ee
where $1+\beta<0$, and $\tau_1<0$.
This generic form of scale factor is
a simple modelling of inflationary expansion,
and the index $ \beta$ is a parameter.
If the inflationary expansion is driven by a scalar field,
then the index $\beta$ is related to
the so-called slow-roll parameters,
$\eta$ and $\epsilon$ \cite{Liddle},
as $\beta=-2+(\eta-3\epsilon)$ .
In this case one usually has  $\beta\le -2$.
In addition, the WMAP5 data combined with Baryon Acoustic Oscillations
and Type Ia supernovae give $n_s =0.960^{+0.014}_{-0.013} $ ($95\%$ CL) \cite{WMAP5}.
Base on the relation $n_s=2\beta+5$ \cite{grishchuk,zhang},
the inflation index $\beta=-2.02$ for $n_s=0.960$.

The reheating stage:
\be \label{reheating}
a(\tau)=a_z|\tau-\tau_p|^{1+\beta_s},\,\,\,\,\tau_1\leq \tau\leq
\tau_s,
\ee
where
$\tau_s$ is the beginning of radiation era and $\tau_p<\tau_1$.
As usual, the model parameter will be taken as $\beta_s= -0.3$ \cite{Miao,Wang}.

The radiation-dominant stage:
\be \label{r}
a(\tau)=a_e(\tau-\tau_e),\,\,\,\,\tau_s\leq \tau\leq \tau_2.
\ee
This is the stage during which the EPT produces PMF.
We use $\tau_{ew}$ to denote the starting time of EPT:
$\tau_s<\tau_{ew} < \tau_2$.
The corresponding energy scale is $T\sim 100$ GeV.
As seen below,
the wave equation of RGW is still homogeneous for $\tau<\tau_{ew}$,
but becomes inhomogeneous for $\tau_{ew} < \tau< \tau_{2}$.

The matter-dominant stage:
\be \label{m}
a(\tau)=a_m(\tau-\tau_m)^2,\,\,\,\,
\tau_2 \leq \tau\leq \tau_E,
\ee
where $\tau_m<\tau_2$ and
$\tau_E$ is the beginning time of the acceleration era.

The accelerating stage:
\be \label{accel}
a(\tau)=l_H|\tau-\tau_a|^{-\gamma},\,\,\,\,\tau_E \leq \tau\leq
\tau_0,
\ee
where $\tau_0$ is the present time and $\tau_0<\tau_a$.
The index $\gamma$ in Eq.(\ref{accel})
depends on the present fractional dark energy $\Omega_\Lambda$.
By fitting with the numerical solution of
the Friedmann equation \cite{zhang,Miao,Wang},
\be \label{Friedmann}
\Big(\frac{a'}{a^2}\Big)^2=\frac{8\pi G}{3} (\rho_\Lambda +\rho_m
+\rho_r),
\ee
where $a'\equiv da/d\tau$,
one can take $\gamma\simeq 1.044$ for $\Omega_{\Lambda}=0.75 $.
The redshift of the start of this stage
depends on the specific models of the dark energy.
For instance, in the $\Lambda CDM$ model with $\Omega_\Lambda =0.72$ and $\Omega_m =0.28$,
it starts at $z\simeq 0.37$;
and in the 3-loop Yang-Mills condensate dark energy model \cite{Wang2}, it starts at $z\simeq 0.5$.

In the above specifications of $a(\tau)$, there are five instances
of time, $\tau_1$, $\tau_s$, $\tau_2$, $\tau_E$, and $\tau_0$, which
separate the different stages. Four of them  are determined by how
much $a(\tau)$ increases over each stage based on the cosmological
considerations.
As in \cite{zhang,Miao,Wang}, we take the following specifications:
$\zeta_1\equiv\frac{a(\tau_s)}{a(\tau_1)}=300$ for the reheating
stage, $\zeta_s\equiv\frac{a(\tau_2)}{a(\tau_s)}=10^{24}$ for the
radiation stage,
$\zeta_2\equiv\frac{a(\tau_E)}{a(\tau_2)}=3454\zeta_E^{-1}$ for the
matter stage, and
$\zeta_E\equiv\frac{a(\tau_0)}{a(\tau_E)}=(\frac{\Omega_\Lambda}{\Omega_m})^{1/3}$
for the present accelerating stage. The remaining time instance is
fixed by an overall normalization \be \label{norm}
|\tau_0-\tau_a|=1. \ee Notice that this convention of normalization
is different from that of Ref. \cite{Durrer2}. There are also 12
constants in the expressions of $a(\tau)$, among which  $\beta$,
$\beta_s$ and $\gamma$ are imposed as the model parameters, for the
inflation, the reheating, and the acceleration, respectively. Based
on the definition of the expansion rate
$H_0=\frac{a'}{a^2}|_{\tau_0}$ of the present universe , one has
$l_H=\gamma/H_0$. Making use of the continuity conditions of
$a(\tau)$ and of $a(\tau)'$ at the four given joining points
$\tau_1$, $\tau_s$, $\tau_2$ and $\tau_E$, all parameters are fixed
as the following:
\begin{eqnarray} \label{aling1}
&&\tau_a-\tau_E=\zeta_E^{\frac{1}{\gamma}},\nonumber\\
&&\tau_E -\tau_m=\frac{2}{\gamma}
                \zeta_E^{\frac{1}{\gamma}},\nonumber\\
&&\tau_2-\tau_m=\frac{2}{\gamma}
      \zeta_2^{-\frac{1}{2}}\zeta_E^{\frac{1}{\gamma}},\nonumber\\
&&\tau_2-\tau_e=\frac{1}{\gamma}\zeta_2^{-\frac{1}{2}}
      \zeta_E^{\frac{1}{\gamma}},              \nonumber\\
&&\tau_s-\tau_e=\frac{1}{\gamma}\zeta_s^{-1}
                 \zeta_2^{-\frac{1}{2}}
                 \zeta_E^{\frac{1}{\gamma}},\nonumber\\
&&\tau_s-\tau_p=\frac{1}{\gamma}(1+\beta_s)
                 \zeta_s^{-1}\zeta_2^{-\frac{1}{2}}
                  \zeta_E^{\frac{1}{\gamma}},\nonumber\\
&&\tau_1-\tau_p=\frac{1}{\gamma}(1+\beta_s)
      \zeta_1^{\frac{-1}{1+\beta_s}}\zeta_s^{-1}
      \zeta_2^{-\frac{1}{2}}\zeta_E^{\frac{1}{\gamma}}, \nonumber\\
&&\tau_1=\frac{1}{\gamma}(1+\beta)
      \zeta_1^{\frac{-1}{1+\beta_s}}\zeta_s^{-1}
      \zeta_2^{-\frac{1}{2}}\zeta_E^{\frac{1}{\gamma}},\nonumber\\
&&\tau_{ew}=2.75\times10^{-11}(\tau_2-\tau_e)-\tau_e,
\end{eqnarray}
and
\begin{eqnarray} \label{aling2}
&&a_m=\frac{l_H}{4}\,\gamma^2\,\zeta_E^{-(1+\frac{2}{\gamma})},\nonumber\\
&&a_e=l_H\,\gamma\,\zeta_2^{-\frac{1}{2}}
    \zeta_E^{-(1+\frac{1}{\gamma})},\nonumber\\
&&a_z=l_H\,\gamma^{1+\beta_s}|1+\beta_s|^{-(1+\beta_s)}
    \zeta_s^{\beta_s}\zeta_2^{\frac{\beta_s-1}{2}}
    \zeta_E^{-(1+\frac{1+\beta_s}{\gamma})},\nonumber\\
&&l_1=l_H\,\gamma^{1+\beta}\,|1+\beta|^{-(1+\beta)}
    \zeta_1^{\frac{\beta-\beta_s}{1+\beta_s}}
    \zeta_s^{\beta}\zeta_2^{\frac{\beta-1}{2}}
    \zeta_E^{-(1+\frac{1+\beta}{\gamma})}.
\end{eqnarray}

In the expanding Universe,
the physical wavelength $\lambda$ is related to the co-moving wave number $k$ by
\be
\lambda\equiv \frac{2\pi a(\tau)}{k},
\ee
and the wave number $k_0$ corresponding to the present Hubble radius is
\be
k_0 = \frac{2\pi a(\tau_0)}{1/H_0}=2\pi \gamma.
\ee
There is another important wave number involved
\be \label{ke}
k_E \equiv \frac{2\pi
a(\tau_E)}{1/H_0}=\frac{k_0}{1+z_E},
\ee
whose corresponding wavelength at the time $\tau_E$ is the Hubble radius $1/H_0$.
In the present universe the physical frequency $\nu$ corresponding to a wave number $k$ is given by
\be \label{12}
\nu = \frac{1}{\lambda}=
\frac{k}{2\pi a (\tau_0)} = \frac{H_0}{2\pi\gamma} k.
\ee

\

\begin{center}
{\bf 3. Analytical solution of RGW}
\end{center}

In the presence of the gravitational waves, the perturbed metric is
\be
ds^2=a^2(\tau)[-d\tau^2+(\delta_{ij}+h_{ij})dx^idx^j],
\ee
where the tensorial perturbation $h_{ij}$
is a $3\times 3$ matrix and is taken to
be transverse and traceless
\begin{eqnarray}
h^i_{\,\,i}=0,&&h_{ij,j}=0.
\end{eqnarray}
The wave equation of RGW is
\be \label{eq:wave}
 \partial_{\nu}( \sqrt{-g}\partial^{\nu} h_{ij}(\tau,{\bf x}) )=0,
\ee
where $g$ is the determinant of the 4-dimensional metric $g_{\mu\nu}$.
One can decompose $h_{ij}$ into the Fourier modes as
\be
\label{planwave}
h_{ij}(\tau,{\bf x})=
\sum_{\sigma}\int\frac{d^3\bf k}{(2\pi)^3}
\epsilon^{\sigma}_{ij}h_{\bf k}^{(\sigma)}(\tau) e^{i\bf{k}\cdot{x}} \, ,
\ee
where $\bf k$ is co-moving wave vector, $\sigma$ denotes polarization states $+,\times$,
and $\epsilon^{(\sigma)}_{ij}$ is the polarization tensor.
Since RGW is isotropic, Eq. (\ref{eq:wave}) reduces to
\be \label{eq:eq}
h_{k}{''}(\tau)
+2\frac{a'(\tau)}{a(\tau)}h_k {'}(\tau)
+k^2h_k(\tau)=0.
\ee
where the co-moving wave number $k$ is the Euclidean length of $\bf k$.
For each polarization, $\times$, $+$,
the equation of $h^{ (\sigma) }_{k}$ is the same,
so the super index $(\sigma)$ can be dropped from $h^{(\sigma)}_{k}$.
Since for all the stages of cosmic expansion the scale factor can be approximately described by a power-law form
\be
a(\tau) \propto \tau^\alpha,
\ee
the solution to Eq. (\ref{eq:eq})
is a linear combination of Bessel function $J_{\nu}$ and Neumann function $N_{\nu}$
\be \label{hom}
h_k(\tau)=x^{(1/2)-\alpha}
 \big[a_1 J_{\alpha-(1/2)}(k \tau)
+a_2   N_{\alpha-(1/2)}(k \tau)\big],
\ee
where the constants $a_1$ and $a_2$ can be determined by the continuity of $h_k$ and of $h'_k$
at the joining points $\tau_1,\tau_s,\tau_2$ and $\tau_E$.
However, as mentioned above, during $\tau_{ew} \leq \tau\le\tau_{2}$,
Eq. (\ref{eq:eq}) will be modified and its solution will be given later.

The inflationary stage has the solution
\be \label{infl}
h_k(\tau)=A_0 l_1^{-1}|\tau|^{-(1/2)-\beta}
\big[ A_1 J_{(1/2)+\beta}(p)
     +A_2 J_{-((1/2)+\beta)}(p) \big],
\,\,\,\, -\infty<\tau\leq \tau_1
\ee
where $p \equiv k\tau$ and
\be
A_1=-\frac{i}{\cos \beta\pi}\sqrt{\frac{\pi}{2}}e^{i\pi\beta/2},
\,\,\,\,\,
A_2=\frac{1}{\cos\beta\pi}\sqrt{\frac{\pi}{2}}e^{-i\pi\beta/2},
\ee
are taken from \cite{grishchuk2},
and thus the so-called \textit{adiabatic vacuum} is achieved:
$\lim_{k\rightarrow \infty}h_k(\tau)\propto e^{-ik\tau} $ in the high frequency limit \cite{parker}.
The constant $A_0$ determined by the initial amplitude of the spectrum is independent of $k$.
For $k\tau\ll 1$, $h_k(\tau)\propto J_{(1/2)+\beta}(x)\propto k^{(1/2)+\beta}$.
As seen below,
this choice will lead to the required scale-invariant initial spectrum in Eq. (\ref{initialspectrum}).

The reheating stage has the solution
\be \label{reh}
h_k(\tau)=t^{-(1/2)-\beta_s}
\Big[B_1J_{(1/2)+\beta_s}(k\,t)
+B_2N_{(1/2)+\beta_s}(k\,t)\Big],
\,\,\,\,\,  \tau_1\leq  \tau\leq\tau_s
\ee
where  $t\equiv \tau-\tau_p$ and
\begin{eqnarray}
&&B_1=-\frac{1}{2}\pi\,t_1^{(3/2)+\beta_s}
\big[kN_{(3/2)+\beta_s}(k\,t_1)h_k(\tau_1)+
N_{(1/2)+\beta_s}(k\,t_1)h_k'(\tau_1)\big],\\
&&B_2=\frac{1}{2}\pi\,t_1^{(3/2)+\beta_s}
\big[kJ_{(3/2)+\beta_s}(k\,t_1)h_k(\tau_1)+
J_{(1/2)+\beta_s}(k\,t_1)h_k'(\tau_1)\big],
\end{eqnarray}
with $t_1\equiv \tau_1-\tau_p$,
and  $h_k(\tau_1 )$ and  $h_k'(\tau_1)$
are the corresponding values from the precedent inflation stage.

The radiation-dominant stage needs to be divided into two parts.
The first part of this stage is before the EPT
when $\tau_s\leq \tau\leq\tau_{ew}$,
the PMF has not been produced yet,
so the wave equation of RGW is still homogeneous with the solution
\begin{equation} \label{r1}
h_k(\tau)=y^{-(1/2)}
\Big[C_1J_{1/2}(k\,y)+C_2N_{1/2}(k\,y)\Big],
\,\,\,\,\, \tau_s \leq \tau\leq\tau_{ew}
\end{equation}
where $y\equiv \tau-\tau_e$ and
\begin{eqnarray}
&&C_1=-\frac{1}{2}\pi\,y_s^{3/2}
\big[kN_{3/2}(k\,y_s)h_k(\tau_s)+
N_{1/2}(k\,y_s)h_k'(\tau_s)\big],\\
&&C_2=\frac{1}{2}\pi\,y_s^{3/2}
\big[kJ_{3/2}(k\,y_s)h_k(\tau_s)+
J_{1/2}(k\,y_s)h_k'(\tau_s)\big],
\end{eqnarray}
with $y_s\equiv \tau_s-\tau_e$, and $h_k(\tau_s )$ and  $h_k'(\tau_s)$ are from the reheating stage.
The second part is from the EPT up to the matter-dominant stage when $\tau_{ew}\leq \tau\leq\tau_2$.
During this period the wave equation of RGW is modified as
\be \label{imheq}
h_{ij}''(\tau)
+2\frac{a'(\tau)}{a(\tau)}h_{ij}'(\tau)
+k^2 h_{ij}(\tau)=16\pi G a^2\Pi_{ij},
\ee
where $\Pi_{ij}$ is an anisotropic term contributed by the PMF.
Note that the righthand side of Eq. (\ref{imheq}) is different from the Eq. (15) in Ref. \cite{Durrer2},
this is because we choose a different normalization of scale factor $|\tau_0-\tau_a|=1$.
As in \cite{Durrer2},
setting $\Pi(k,\tau)=(\frac{a_{0}}{a})^{2}f(k,\tau)\tilde\Pi(k)$
(where $a_{0}$ is today's scale factor and $\tilde\Pi(k)$ is a time independent random variable
with power spectrum $\langle|\tilde\Pi(k)|^2\rangle =1$),
Eq. (\ref{imheq}) can be reduced to
\be \label{imheq2}
h_{k}''(\tau)+2\frac{a'(\tau)}{a(\tau)}h_k'(\tau)
+k^2h_k(\tau)=16\pi G a_{0}^2 f(k,\tau).
\ee
For the PMF produced by EPT,
the source is a white noise independent of $k$ \cite{Durrer2},
and the key function $f$ can be written as (See Appendix A for detailed derivations)
\be  \label{f}
f(\tau)=\frac{B_{\lambda}^{2}(\lambda / \sqrt{2})^{n+(3/2)}}{2\pi^{9/4}\Gamma
        (\frac{n+3}{2})}k_{c}(\tau)^{n+(3/2)}.
\ee
$B_{\lambda}$ is the amplitude of PMF,
$\lambda$ is the length scale
on which cosmic magnetic fields have been observed.
For the PMF produced by EPT, the spectral index $n=2$,
and $k_{c}=1/\eta_{in}$ with $\eta_{in} \simeq 4\times10^{4}sec$ is the corresponding cutoff scale \cite{Durrer2}.
By using the theorem of Wronskian,
Eq. (\ref{imheq2}) can be analytically solved (See Appendix B for details),
and then the effects of PMF can be included into our calculation of RGW.

The matter-dominant stage has the solution
\begin{equation}  \label{matter}
h_k(\tau)=\varsigma^{-(3/2)}
\Big[D_1J_{3/2}(k\,\varsigma)+D_2N_{3/2}(k\,\varsigma)\Big],
\,\,\,\, \tau_2\leq \tau\leq\tau_E
\end{equation}
where $\varsigma \equiv \tau-\tau_m$ and
\begin{eqnarray}
&&D_1=-\frac{1}{2}\pi\,\varsigma_2^{5/2}
\big[kN_{5/2}(k\,\varsigma_2)h_k(\tau_2)+
N_{3/2}(k\,\varsigma_2)h_k'(\tau_2)\big],\\
&&D_2=\frac{1}{2}\pi\,\varsigma_2^{5/2}
\big[kJ_{5/2}(k\,\varsigma_2)h_k(\tau_2)+
J_{3/2}(k\,\varsigma_2)h_k'(\tau_2)\big],
\end{eqnarray}
with $\varsigma_2\equiv \tau_2-\tau_m$.
In the expressions of $D_1$ and $D_2$,
the mode functions $h_k(\tau_2)$ and $h'_k(\tau_2)$ are also from the precedent stage.

The accelerating stage has the solution
\be \label{acc}
h_k(\tau)=s^{(1/2)+\gamma}
\Big[E_1J_{-(1/2)-\gamma}(k\,s)
+E_2N_{-(1/2)-\gamma}(k\,s)\Big],
\,\,\,\,\, \tau_E\leq\tau\leq\tau_0
\ee
where $s\equiv \tau-\tau_a$  and
\begin{eqnarray}
&&E_1=-\frac{1}{2}\pi\,s_E^{(1/2)-\gamma}
\big[kN_{(1/2)-\gamma}(k\,s_E)h_k(\tau_E)+
N_{-(1/2)-\gamma}(k\,s_E)h_k'(\tau_E)\big],\\
&&E_2=\frac{1}{2}\pi\,s_E^{(1/2)-\gamma}
\big[kJ_{(1/2)-\gamma}(k\,s_E)h_k(\tau_E)+
J_{-(1/2)-\gamma}(k\,s_E)h_k'(\tau_E)\big],
\end{eqnarray}
with  $s_E\equiv \tau_E-\tau_a$.
So far, the explicit solution of $h_k(\tau)$
has been obtained for all the expansion stages,
from Eq. (\ref{infl}) through Eq. (\ref{acc}).

\

\begin{center}
 {\bf 4. Theoretical Predictions of RGW}
\end{center}

There are three different spectra that are commonly used to describe the stochastic GW background:
the tensor power spectrum $\Delta^2(k,\tau)$, the gravitational wave spectrum $h(k,\tau)$,
and the energy density spectrum $\Omega_{gw}(k,\tau)$.
In the early universe, the RGW is usually characterized by $\Delta^2(k,\tau)$;
on the other hand, the present-day RGW is usually characterized by $h(k,\tau)$ or by $\Omega_{gw}(k,\tau)$.
In particular, as pointed by Grishchuk \cite{grishchuk1},
the spectrum of RGW $h(k,\tau)$ can be directly compared with the direct measurements of Laser interferometer GW detection,
and is defined by the following equation
\cite{grishchuk1}:
\be \label{gr1}
\int_0^{\infty}h^2(k,\tau)\frac{dk}{k}\equiv\langle0| h^{ij}({\bf
x},\tau)h_{ij}({\bf x},\tau)|0\rangle,
\ee
where the right-hand side is the expectation value of the $h^{ij}h_{ij}$.
Calculation yields the spectrum at present
\be
\label{33} h(k,\tau_0) =
\frac{2}{\pi}k^{3/2}|h_k(\tau_0)|.
\ee
One of the most important properties of the inflation is that the initial spectrum of RGW
at the time $\tau_i$ of the horizon-crossing during the inflation is nearly scale-invariant \cite{grishchuk1}:
\be
\label{initialspectrum} h(k,\tau_i)
=A\Big(\frac{k}{k_0}\Big)^{2+\beta},
\ee
where constant $A$ is directly proportional to $A_{0}$ in Eq. (\ref{infl}),
and will be fixed by the observed CMB anisotropies in practice.
Since the observed CMB anisotropies \cite{Spergel} are $\Delta T/T \simeq 0.37\times 10^{-5}$ at $l\sim 10$,
as in Refs. \cite{Miao,Wang}, we take the normalization of the spectrum
\be
\label{norml}
h(k_E,\tau_0)=0.37\times10^{-5}r^{\frac{1}{2}},
\ee
where $k_E$ is
defined in Eq. (\ref{ke}), and its physical frequency being $\nu_E =
k_E/2\pi a(\tau_0)=H_0/(1+z_E)\sim 1 \times 10^{-18}$ Hz. The
tensor/scalar ratio $r$ can be related to the slow roll parameter
$\epsilon$ in the scalar inflationary model as $r=16\epsilon$
\cite{Liddle}. However, the value of $r$ is model-dependent, and
frequency-dependent \cite{zhao2,baskaran}. This has long been known
to be a notoriously thorny issue \cite{seljak2}. In our treatment,
for simplicity, $r$ is only taken as a constant parameter for
normalization of RGW. The WMAP5 data combined with Baryon Acoustic
Oscillations and Type Ia supernovae  give $r< 0.20 $ ($95\%$ CL)
\cite{WMAP5}. For concreteness, we take $r=0.20$.

Next, we shall focus on how the theoretical prediction of RGW is related to the GW experiment.
As pointed by \cite{Maggiore},
\be \label{key1}
\langle0| h^{ij}({\bf x},\tau_{0})h_{ij}({\bf x},\tau_{0})|0\rangle
=4\int_0^{\infty}\nu S_h(\nu)\frac{d\nu}{\nu},
\ee
where $S_h(\nu)$ is the one-sided spectral density of RGW.
Base on Eqs. (\ref{gr1}), (\ref{key1}),
and using Eq. (\ref{12}) to change $dk/k$ into $d\nu/ \nu$, one gets
\be \label{oss1}
\sqrt{S_h(\nu)}=\frac{h(\nu,\tau_0)}{2\sqrt{\nu}}.
\ee
The noise level of the detector is measured by the strain sensitivity $h_{n}$,
defined as
\be \label{oss2}
h_{n}\equiv\sqrt{S_n(\nu)},
\ee
where $S_n(\nu)$ is the one-sided spectral density of the detector's noise.
A stochastic GW background will be observable at a frequency $\nu$ if \cite{Maggiore}
\be \label{oss3}
S_h(\nu)>S_n(\nu)/F,
\ee
where $F$ is the angular efficiency factor determined by the geometry of the detector.
For interferometers, $F=2/5$ is always satisfied \cite{Maggiore}.
Therefore, to compare the theoretical model of RGW with the latest LIGO data,
one needs to calculate
\be \label{result}
\frac{h(\nu,\tau_0)}{\sqrt{10\nu}}.
\ee
In principle, since all the RGW solutions of the 6 cosmic stages are analytic,
the analytic expression of $h(\nu,\tau_0)$ can be written down.
However, since this expression is too complex, we only give the numerical results in this work.

\

\begin{center}
 {\bf 5. Results and Discussions}
\end{center}

In the following, by comparing our theoretical results with LIGO S5 data, we will give the constraints on the amplitude of PMF.

\begin{figure}
\centerline{\includegraphics[width=8cm]{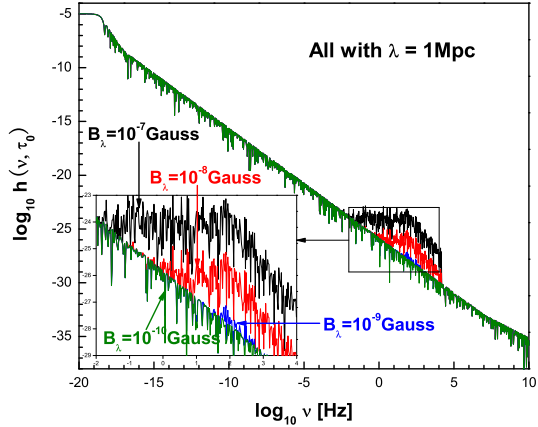}}
\caption{\label{fig1}
The spectrum $h(\nu,\tau_0)$ of RGW for various amplitudes $B_{\lambda}$ of PMF.
PMF will not markedly affect $h(\nu,\tau_0)$, unless its amplitude is stronger than $B_{\lambda} \geq 10^{-9}$ Gauss.}
\end{figure}

In Figure \ref{fig1}, we plot the spectrum $h(\nu,\tau_0)$ of RGW for various amplitudes $B_{\lambda}$ of PMF,
where the scale of galactic cluster $\lambda=1$ Mpc is adopted.
PMF yields a level spectrum of RGW in the range $(10^{-2}-10^{4})$ Hz,
which covers the bands of operation of LIGO.
For a larger amplitude $B_{\lambda}$,
this level spectrum of RGW will also be higher.
It is seen that PMF will not markedly affect $h(\nu,\tau_0)$,
unless its amplitude is stronger than $B_{\lambda} \geq 10^{-9}$ Gauss.

\begin{figure}
\centerline{\includegraphics[width=8cm]{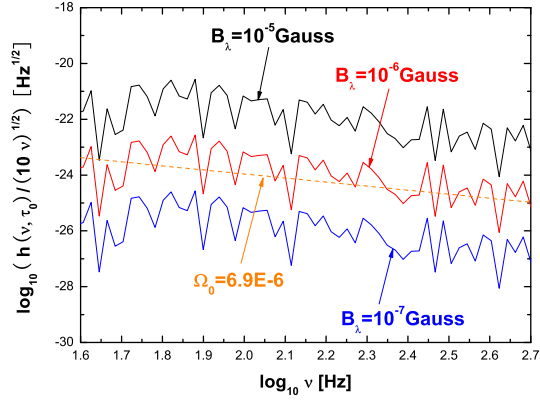}}
\caption{\label{fig2}
Comparison of the theoretical predictions of RGW with the latest LIGO S5 data,
where the scale of galactic cluster $\lambda=1$ Mpc is adopted in this comparison.}
\end{figure}

Figure \ref{fig2} is a comparison of the theoretical predictions of RGW with the latest LIGO S5 data \cite{ligo}.
Again, a scale of galactic cluster $\lambda=1$ Mpc is adopted in this comparison.
The three fluctuating curves are the RGW's theoretical prediction of
$B_{\lambda}= 10^{-5}$ Gauss, $B_{\lambda}= 10^{-6}$ Gauss, and $B_{\lambda}= 10^{-7}$ Gauss, respectively.
The dashed line is the latest lower limit given by the LIGO S5 run \cite{ligo}.
Here the vertical axis is the strain sensitivity, given by Eq. (\ref{result}).
We find that for the scale of galactic cluster $\lambda=1$ Mpc,
the amplitude of PMF has to be weaker than $B_{\lambda} \leq 4\times 10^{-7}$ Gauss.

\begin{figure}
\centerline{\includegraphics[width=8cm]{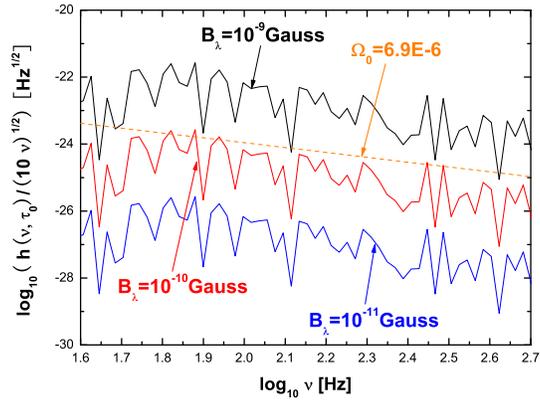}}
\caption{\label{fig3}
Comparison of the theoretical predictions of RGW with the latest LIGO S5 data,
where the scale of supercluster $\lambda=100$ Mpc is adopted in this comparison.}
\end{figure}

Now, let us turn to the case of PMF with a scale of supercluster.
Figure \ref{fig3} is a comparison of the theoretical predictions of RGW with the latest LIGO S5 data,
where the scale of supercluster $\lambda=100$ Mpc is adopted.
The three fluctuating curves are the RGW's theoretical prediction of
$B_{\lambda}= 10^{-9}$ Gauss, $B_{\lambda}= 10^{-10}$ Gauss, and $B_{\lambda}= 10^{-11}$ Gauss, respectively.
The dashed line is the latest lower limit given by the LIGO S5 run \cite{ligo}.
It is found that for the case of $\lambda=100$ Mpc,
the amplitude of PMF has to be weaker than $B_{\lambda} \leq 9\times 10^{-11}$ Gauss.

In this paper, we present new PMF limits by using LIGO S5 data.
Compared with the constraints from the nucleosynthesis limit \cite{Durrer2},
our results are much closer to that from the WMAP5 $B$-mode polarization limit \cite{Giovannini2}.
Although our constraints are not as strong as those from CMB observations,
future laser interferometer GW detectors,
such as advanced LIGO \cite{ligo2}, BBO \cite{BBO} and DECIGO \cite{DECIGO},
will be able to put much tighter constraints on the PMF.
In this manner, current and future GW detectors have potential
to make interesting contributions to the study of the PMF.

\

ACKNOWLEDGMENT:
We are grateful the referees for valuable suggestions.
We thank Prof. Yang Zhang, Prof. Pedro Ferreira, Dr. Tianyang Xia, and Dr. Linqing Wen for helpful discussions and kind help.
We are also grateful Prof. Miao Li for a careful reading of the manuscript.
The author was supported by a graduate fund of USTC.


\

\begin{center}
{\bf Appendix A: Detailed derivations of the key function $f$}
\end{center}

As in Ref. \cite{Durrer2},
we shall only consider the magnetic fields on sufficiently large scales, and ignore the impacts come from small scales.
We model today's magnetic field ${\bf B}_0({\bf x})$ as a statistically homogeneous and isotropic random field.
The transversal nature of magnetic field ${\bf B}$ then leads to
\be \label{A1}
\langle B^i({\bf k})B^{*j}({\bf q})\rangle =
\delta^3({\bf k-q})(\delta^{ij}-\hat{k}^i\hat{k}^j)B^2(k),
\ee
where $\bf k$ and $\bf q$ are co-moving wave vectors, $\hat{k} ={\bf k}/k$,
and $k =\sqrt{\sum_i(k^i)^2}$ is the Euclidean length of $\bf k$.
We use the following Fourier transform conventions
\be \label{A2}
B^j({\bf k})=\sum_{{\bf k}}
\exp(i{\bf x\cdot k})B_0^j({\bf x}), ~~
B_0^j({\bf x})=\frac{1}{V}\int d^3k \exp(-i{\bf x\cdot k})
B^{j}({\bf k}),
\ee
where $V=\int d^3r \exp(-\frac{r^{2}}{2\lambda^{2}})$ is the normalization volume
($\lambda$ is the length scale of cosmic magnetic fields).
On sufficiently large scales,
$B^{2}({\bf x})=\Big(\frac{a_{0}}{a}\Big)^{4}B^{2}_{0}({\bf x})$,
$B_{i}({\bf x})=\Big(\frac{a_{0}}{a}\Big)B_{0i}({\bf x})$,
and $B^{i}({\bf x})=\Big(\frac{a_{0}}{a}\Big)^{3}B^{i}_{0}({\bf x})$.
If $\bf B$ is generated by a {\it causal} mechanism,
it is uncorrelated on super horizon scales,
\be \label{A3}
\langle B^i({\bf x},\tau)B^j({\bf x}',\tau)\rangle=0
~~~{\rm for }~~~ |{\bf x}-{\bf x}'| > 2\tau.
\ee
Notice that the universe is in a stage of standard Friedman expansion when the PMF is produced,
so that the co-moving causal horizon size is about $\tau$.
According to Eq. (\ref{A3}),
$ \langle B^i({\bf x},\tau)B^j({\bf x}',\tau)\rangle$ is a function
with compact support and hence its Fourier transform is analytic,
i.e.
\be \label{A4}
\langle B^i({\bf k})B^{*j}({\bf k})\rangle
          \equiv (\delta^{ij}-\hat{k}^i\hat{k}^j)B^2(k)
\ee
is analytic in {\bf k}.
The Maxwell stress tensor $T^{ij}$ of a magnetic field in real space is given by
\be \label{A5}
T^{ij}({\bf x},\tau)={1\over 4\pi}\Big[B^i({\bf x},\tau)B^j({\bf x},\tau)
-{1\over 2} g^{ij}({\bf x},\tau)B_l({\bf x},\tau)B^l({\bf x},\tau)\Big].
\ee
In Fourier space, making use of Eq. (\ref{A2}) and the scaling of the magnetic field with time, we have
\be \label{A6}
T^{ij}({\bf k},\tau)=\frac{1}{4\pi}\Big(\frac{a_{0}}{a}\Big)^{6}\sum_{{\bf q}}
\Big[B^{i}({\bf  q})B^{j}({\bf k-q}) -{1\over 2}B_{l}({\bf q})B^{l}({\bf k-q})\delta^{ij}\Big].
\ee
$\Pi^{ij}(\bf k,\tau)$ is the transverse traceless component of $T^{ij}(\bf k,\tau)$
and sources the RGW, defined as
\be \label{A7}
\Pi^{ij}({\bf k},\tau)\equiv(P^{i}_{a}P^{j}_{b}-{1\over 2}P^{ij}P_{ab})T^{ab}({\bf k},\tau),
\ee
where $P_{ij} = \delta_{ij}-\hat{k}_i\hat{k}_j$ is the projector onto the component of a vector transverse to $\bf k$,
and $P^{i}_{a}P^{j}_{b}$ projects onto the transverse component of a tensor.
Here we give the details of the calculation of its correlation function $\langle\Pi^{ij}({\bf k},\tau)\Pi^{*lm}({\bf k'},\tau)\rangle$,
which will be used to compute the induced GW.
Defining the projector
${\mathcal P}^{ij}_{ab}=P^{i}_{a}P^{j}_{b}-{1\over 2}P^{ij}P_{ab}$,
we have
\be \label{A8}
\langle\Pi^{ij}({\bf k},\tau)\Pi^{*lm}({\bf k'},\tau)\rangle
          ={\mathcal P}^{ij}_{~~ab}{\mathcal P}^{lm}_{~~cd}
\langle T^{ab}({\bf k},\tau)T^{*cd}({\bf k'},\tau)\rangle.
\ee
Note that up to a trace,
which anyway vanishes in the projection (\ref{A8}),
$T^{ab}(\bf k,\tau)$ is just given by
\be \label{A9}
\Delta^{ab}({\bf k},\tau) \equiv \frac{1}{4 \pi}\Big(\frac{a_{0}}{a}\Big)^{6}
\sum_{\bf q} B^{a}({\bf q}) B^{b}({\bf k-q}).
\ee
Therefore, we can get
\be \label{A10}
\langle\Pi^{ij}({\bf k},\tau)\Pi^{*lm}({\bf k'},\tau)\rangle
={\mathcal P}^{ij}_{~~ab}{\mathcal P}^{lm}_{~~cd}
\langle\Delta^{ab}(\bf k,\tau) \Delta^{*cd}(\bf k',\tau)\rangle.
\ee
Assuming the random magnetic field be Gaussian,
the Wick's theorem for Gaussian fields can be applied here.
Making use of Eq. (\ref{A1}), Eq. (\ref{A9}), and the reality condition $B^{*a}(k)=B^a(-k)$,
products of four magnetic fields can be reduced as
\begin{eqnarray} \label{A11}
&& \langle B^a({\bf q})B^{b}({\bf k-q}) B^{*c}({\bf p})B^{*d}({\bf k'-p})\rangle \nonumber \\&&
= \delta_{\bf q,\bf q-k}(\delta^{ab}-\hat{q}^{a}\hat{q}^{b})B^{2}(q) \cdot \delta_{\bf -p,\bf k'-p}(\delta^{cd}-\hat{q}^{c}\hat{q}^{d})B^{2}(-p)
\nonumber
\\&& + \delta_{\bf q,\bf p}(\delta^{ac}-\hat{q}^{a}\hat{q}^{c})B^{2}(q) \cdot \delta_{\bf k-q,\bf k'-p}(\delta^{bd}-\widehat{k-q}^{b}\widehat{k-q}^{d})B^{2}(|\bf q-k|)
\nonumber
\\&& + \delta_{\bf q,\bf k'-p}(\delta^{ad}-\hat{q}^{a}\hat{q}^{d})B^{2}(q) \cdot \delta_{\bf k-q,\bf p}(\delta^{bc}-\widehat{k-q}^{b}\widehat{k-q}^{c})B^{2}(|\bf q-k|).
\end{eqnarray}
The first term only contributes an unimportant constant and then can be disregarded.
Making use of Eqs. (\ref{A9}), (\ref{A10}), (\ref{A11}),
and setting ${\mathcal P}_{abcd}={\mathcal P}_{ijab}{\mathcal P}^{ij}_{cd} ={\mathcal P}_{abij}{\mathcal P}_{cd}^{ij}$,
one can obtain
\begin{eqnarray} \label{A12}
&& \langle\Pi_{ij}({\bf k},\tau)\Pi^{*ij}({\bf k'},\tau)\rangle
=(\frac{a}{a_{0}})^{4}{\mathcal P}_{abcd}\langle\Delta^{ab}({\bf
k},\tau)\Delta^{*cd}({\bf k'},\tau)\rangle \nonumber \\&&
=\frac{(a_{0}/a)^{8}}{16\pi^{2}}\delta_{\bf k,\bf k'} \sum_{{\bf
q}}B^{2}(q)B^{2}(|\bf k-q|)(1+2\gamma^{2}+\gamma^{2}\beta^{2}),
\end{eqnarray}
where $\gamma={\bf \hat k}\cdot{\bf \hat q}$,
and $\beta = {\bf \hat k}\cdot { \widehat{\bf k-q}}$.
As in Ref. \cite{Durrer2}, setting
\be \label{A13}
\langle\Pi_{ij}({\bf k},\tau)\Pi^{*ij}({\bf k'},\tau)\rangle
=4(\frac{a_{0}}{a})^{8}f^{2}(k)\delta_{\bf k,\bf k'},
\ee
one can get
\be \label{A14}
f^{2}(k)=(\frac{1}{64\pi^{2}})
\sum_{{\bf q}}B^{2}(q)B^{2}(|\bf k-q|)(1+2\gamma^{2}+\gamma^{2}\beta^{2}).
\ee
In a coordinate system where $\bf k$ is parallel to the $z$-axis,  $\Pi_{ij}$ has the form
\[
\left(\Pi_{ij}\right) = \left(\begin{array}{ccc}
   \Pi_+ & \Pi_\times & 0 \\
  \Pi_\times & -\Pi_+ & 0 \\
   0 & 0 & 0 \end{array} \right)~,
\]
where $+$ and $\times$ denote the polarization states of GW.
Base on Eq. (\ref{A13}), the statistical isotropy gives
\be \label{add}
\langle|\Pi_+|^2\rangle =
\langle|\Pi_\times|^2\rangle =(\frac{a_{0}}{a})^{4}f^2.
\ee
To continue, we have to specify $B^2(k)$.
Base on Eq. (\ref{A2}), one gets
\be \label{A15}
\langle B_{0}({\bf k})B_{0}({\bf x+r})\rangle
=\frac{1}{V}\int d^{3}x\sum_{{\bf k}}\sum_{{\bf k'}}
e^{-i({\bf k+\bf k'})\cdot \bf x}
e^{-i{\bf k'\cdot r}}B({\bf k})B({\bf k'}),
\ee
and the normalization is
\be \label{A16}
B_{\lambda}^{2}=\frac{1}{V}\int d^{3}r
\langle B_{0}({\bf k})B_{0}({\bf x+r})\rangle e^{-\frac{r^{2}}{2\lambda^{2}}}.
\ee
As in Ref. \cite{Durrer2}, we assume that $B^2(k)$ can be approximated by a simple power law,
i.e. $B^2(k)=C k^n$ (C is a constant).
Base on Eq. (\ref{A16}), a tedious but straight forward computation gives
\be  \label{A17}
C=\frac{B_{\lambda}^{2}(\lambda / \sqrt{2})^{n}}{2\pi^{5/2}\Gamma(\frac{n+3}{2})},
\ee
and thus
\be \label{A18}
B^2(k)=\frac{B_{\lambda}^{2}(\lambda /
\sqrt{2})^{n}}{2\pi^{5/2}\Gamma(\frac{n+3}{2})}k^n.
\ee
Making use of Eqs. (\ref{A14}) and (\ref{A18}), one gets
\be \label{A19}
f^{2}(k)=\frac{B_{\lambda}^{4}(\lambda / \sqrt{2})^{2n+3}}{32\pi^{11/2}\Gamma^{2}
(\frac{n+3}{2})}
\int d^{3}q (q)^{n}(|\bf k-q|)^{n}(1+2\gamma^{2}+\gamma^{2}\beta^{2}).
\ee
Now let us focus on the PMF produced by EPT.
As mentioned above, for this case,
and the source is a white noise independent of $k$.
Utilizing the integration result of Ref. \cite{Durrer2}, finally we get
\be \label{A20}
f^{2}=\frac{B_{\lambda}^{4}(\lambda / \sqrt{2})^{2n+3}}{4\pi^{9/2}\Gamma^{2}
(\frac{n+3}{2})}k_{c}^{2n+3},
\ee
where the spectral index $n=2$, and the corresponding cutoff scale is
$k_{c}=1/\eta_{in}$ with $\eta_{in} \simeq 4\times10^{4}sec$ \cite{Durrer2}.
This leads to the expression of key function $f$ we need in Eq. (\ref{imheq2}).
Note that the expression of $f$ has subtle difference with that in \cite{Durrer2},
this is because we use different conventions of dimension.

\begin{center}
{\bf Appendix B: Exact Analytic Solution of Differential Equation (\ref{imheq2})}
\end{center}

As mentioned above, to obtain the solution of RGW during the period $\tau_{ew}\leq \tau\leq\tau_2$,
one must solve the differential equation (\ref{imheq2}).
Follow \cite{Durrer2}, we can write the mode function as
\be \label{B1}
h_k(\tau) = h_k(\tau_{ew})\chi(u),
\ee
where $h_k(\tau_{ew})$ is given by Eq. (\ref{r1}) evaluated at $\tau_{ew}$,
and $\chi(u)$ satisfies the following differential equation
\be\label{B2}
\chi''(u)+\frac{2}{u}\chi'(u)+\chi(u)=\frac{s(k,\tau)}{k^{2}h_k(\tau_{ew})},
\ee
with $u\equiv k\tau$, and $s(k,\tau)=16\pi G a_{0}^2 f(k,\tau)$.
The homogeneous solution of Eq.(\ref{B2}) is
\be\label{B3}
\chi_{1}(u)=W_{1}\frac{\cos(u)}{u}+W_{2}\frac{\sin(u)}{u},
\ee
where the coefficients
\begin{eqnarray}
&&W_1=u_{ew}\cos(u_{ew})+(W u_{ew}-1)\sin(u_{ew}),\\
&&W_2=u_{ew}\sin(u_{ew})-(W u_{ew}-1)\cos(u_{ew}),
\end{eqnarray}
are fixed by the continuity condition
of $h_{k}$ and $h'_{k}$ at the time instance $\tau=\tau_{ew}$,
with $u_{ew}\equiv k\tau_{ew}$,
and
\be\label{B6}
W=\frac{B_{1}J_{3/2}[k(\tau_{ew}-\tau_{e})]+B_{2}N_{3/2}[k(\tau_{ew}-\tau_{e})]}
{B_{1}J_{1/2}[k(\tau_{ew}-\tau_{e})]+B_{2}N_{1/2}[k(\tau_{ew}-\tau_{e})]}.
\ee
Using the theorem of Wronskian,
one can get a special solution of Eq.(\ref{B2})
\be\label{B7}
\chi_{2}(u)=\int_{u_{ew}}^{u}\frac{U(\sin u\cos U-\sin U\cos u)}{k^{2}u h_k(\tau_{ew})}s(k,U)d U.
\ee
Therefore, the general solution of Eq.(\ref{B2}) can be written as
\be\label{B8}
\chi(u)=\chi_{1}(u)+\chi_{2}(u),
\ee
and thus the solution of $h_k(\tau)$ at the period $\tau_{ew}\leq \tau\leq\tau_2$ is in hand.



\small
\begin{thebibliography}{80}

\bibitem[*]{email}{e-mail: swang@mail.ustc.edu.cn}

\bibitem{Kronberg} P.P. Kronberg, Rep. Prog. Phys. {\bf 57}, 57 (1994);
T.E. Clarke, P.P. Kronberg, and H. Boehringer, ApJL. {\bf 547}, 111 (2001);
Y. Xu, P.P. Kronberg, S. Habib, and Q.W. Dufton, ApJ. 637, 19 (2006).

\bibitem{Widrow} L.M. Widrow, Rev. Mod. Phys. {\bf 74}, 775 (2003);
M. Giovannini, Int. J. Mod. Phys. D {\bf 13}, 391 (2004);
J.D. Barrow, R. Maartens, and C.G. Tsagas, Phys. Rep. {\bf 449}, 131 (2007).

\bibitem{Turner} M.S. Turner and L.M. Widrow, Phys. Rev. D {\bf 37}, 2743 (1988).

\bibitem{Ratra} B. Ratra, ApJL. {\bf 391}, 1 (1992).

\bibitem{Ohta} K. Bamba, N. Ohta, and S. Tsujikawa, Phys. Rev. D {\bf 78}, 043524 (2008).


\bibitem{Baym} G. Baym, D. Bodeker and L. McLerran, Phys. Rev. D {\bf 53}, 662 (1996).


\bibitem{Sigl} G. Sigl, A.V. Olinto and K. Jedamzik, Phys. Rev. D {\bf 55}, 4582 (1997).


\bibitem{KKS} T. Kahniashvili, L. Kisslinger and T. Stevens, arXiv:0905.0643.


\bibitem{Kosowsky1} A. Kosowsky and A. Loeb, ApJ. {\bf 469}, 1 (1996).

\bibitem{Giovannini1} M. Giovannini, Phys. Rev. D {\bf 56}, 3198 (1997).

\bibitem{Barrow} J.D. Barrow, P.G. Ferreira and J. Silk, Phys. Rev. Lett. {\bf 78}, 3610 (1997).

\bibitem{Durrer1} R. Durrer, T. Kahniashvili and A. Yates, Phys. Rev. D {\bf 58}, 123004 (1998);
R. Durrer, P.G. Ferreira and T. Kahniashvili, Phys. Rev. D {\bf 61}, 043001 (2000).

\bibitem{Kosowsky2} A. Kosowsky, T. Kahniashvili, G. Lavrelashvili and B. Ratra, Phys. Rev. D {\bf 71}, 043006 (2005).

\bibitem{KLR} T. Kahniashvili, G. Lavrelashvili and B. Ratra, Phys. Rev. D {\bf 78}, 063012 (2008).

\bibitem{Kahniashvili} T. Kahniashvili, Y. Maravin and A. Kosowsky, arXiv:0806.1876.

\bibitem{Kristiansen} J. R. Kristiansen and P. G. Ferreira, Phys. Rev. D {\bf 77}, 123004 (2008).

\bibitem{Mack} A. Mack, T. Kahniashvili, and A. Kosowsky, Phys. Rev. D {\bf 65} 123004 (2002).

\bibitem{Tsagas} C.G. Tsagas, Class. Quant. Grav.{\bf 19}, 3709 (2002).

\bibitem{Durrer2}
C. Caprini and R. Durrer, Phys. Rev. D {\bf 65}, 023517 (2001); Phys. Rev. D {\bf 74}, 063521 (2006).

\bibitem{Giovannini2} M. Giovannini and K.E. Kunze, Phys. Rev. D {\bf 78}, 023010 (2008).

\bibitem{Allen} B. Allen and J.D. Romano, Phys. Rev. D{\bf 59}, 102001 (1999).

\bibitem{Abbott1} B. Abbott et al., ApJ. 659, 918 (2007); Phys. Rev. D {\bf 76}, 082003 (2007).


\bibitem{Abbott2} B. Abbott et al., Phys. Rev. D {\bf 76}, 022001 (2007).


\bibitem{ligo} The LIGO Scientific Collaboration and The Virgo Collaboration,  Nature {\bf 460}, 990 (2009)


\bibitem{grishchuk1} L.P. Grishchuk, Lect. Notes. Phys. {\bf 562}, 167 (2001).


\bibitem{Liddle} A.R. Liddle and D.H. Lyth, Phys. Lett. B {\bf 291}, 391 (2006).


\bibitem{WMAP5} G. Hinshaw, et al, ApJS, {\bf 180}, 225 (2009);
E. Komatsu,, et al, ApJS, {\bf 180}, 330 (2009).


\bibitem{grishchuk} L.P. Grishchuk, Sov. Phys. JETP {\bf 40}, 409 (1975); Phys. Usp. {\bf 44}, 1 (2001).


\bibitem{zhang} Y. Zhang, et al, Class. Quant. Grav. {\bf 22}, 1817 (2005);
Y. Zhang, et al, Class. Quant. Grav. {\bf 23}, 3783 (2006).


\bibitem{Miao} H.X. Miao and Y. Zhang, Phys. Rev. D {\bf 75}, 104009 (2007).


\bibitem{Wang} S. Wang, Y. Zhang, T.Y. Xia, H.X. Miao, Phys. Rev. D {\bf 77}, 104016 (2008).


\bibitem{Wang2} S. Wang, Y. Zhang and T.Y. Xia, JCAP {\bf 10}, 037 (2008);
S. Wang and Y. Zhang, Phys. Lett. B {\bf 669}, 201 (2008).


\bibitem{grishchuk2} L.P. Grishchuk, Phys. Rev. D {\bf 48}, 3513 (1993);
Class. Quant. Grav. {\bf 14}, 1445 (1997).



\bibitem{parker} L. Parker, Phys. Rev. {\bf 183}, 1057 (1969).


\bibitem{Spergel} D.N. Spergel, et al,  ApJS {\bf 148}, 175 (2003).

         D.N. Spergel, et.al. ApJS {\bf 170}, 377 (2007).

\bibitem{zhao2} W. Zhao and Y. Zhang, Phys. Rev. D {\bf 74}, 083006 (2006).

\bibitem{baskaran} D. Baskaran, L.P. Grishchuk, and A.G. Polnarev, Phys. Rev. D {\bf 74}, 083008 (2006);
A.G. Polnarev, N.J. Miller and B.G. Keating, MNRAS, {\bf 386}, 1053, (2008)


\bibitem{seljak2} U. Seljak, et.al., Phys. Rev. D {\bf 71} 103515 (2005);
A. Cooray, P.S. Corasaniti, T. Giannantonio and A. Melchiorri, Phys. Rev. D {\bf 72}, 023514 (2005);
A. Linde, V. Mukhanov and M. Sasaki, JCAP {\bf 0510}, 002 (2005);
V. Mukhanov and A.Vikman, JCAP {\bf 0602}, 004 (2006).
T.L. Smith, M. Kamionkowski and A. Cooray, Phys. Rev. D {\bf 73}, 023504 (2006);
T.L. Smith, M. Kamionkowski and A. Cooray, Phys. Rev. D {\bf 78}, 083525 (2008).

\bibitem{Maggiore} M. Maggiore, Phys. Rep. {\bf 331}, 283 (2000).


\bibitem{ligo2} http://www.ligo.caltech.edu/advLIGO.


\bibitem{BBO} http://universe.nasa.gov/program/bbo.html



\bibitem{DECIGO} N. Seto, S. Kawamura, and T. Nakamura, Phys. Rev. Lett. {\bf 87}, 221103 (2001).


\end{thebibliography}
\end{document}